\documentclass[useAMS,usenatbib]{mn2e}

\usepackage{graphicx}
\usepackage{rotating}
\usepackage{longtable}
\usepackage{lscape}

\def\Mbh{$M_{\rm BH}$}
\def\Lhost{$L_{\rm host}$}

\def\Civ{C\,{\sc iv}}
\def\Mgii{Mg\,{\sc ii}}
\def\FeII{Fe\,{\sc ii}}

\def\lsim{\mathrel{\rlap{\lower 3pt \hbox{$\sim$}} \raise 2.0pt \hbox{$<$}}}
\def\gsim{\mathrel{\rlap{\lower 3pt \hbox{$\sim$}} \raise 2.0pt \hbox{$>$}}}

\title[Q2225-403A,B]{A new apparent quasar pair: Q2225-403A,B}
\author[Decarli et al.]{R. Decarli$^{1}$\thanks{E-mail:roberto.decarli@mib.infn.it},
A. Treves$^{1}$, R. Falomo$^{2}$\\
$^{1}$ Universit\`{a} degli Studi dell'Insubria, via Valleggio 11,
          22100 Como, Italy \\
$^{2}$ INAF - Osservatorio Astronomico di Padova, Vicolo dell'Osservatorio 5,
             35122, Padova, Italy\\
}
\begin{document}

\date{ }

\pagerange{\pageref{firstpage}--\pageref{lastpage}} \pubyear{2009}

\maketitle

\label{firstpage}

\begin{abstract}
We report the serendipitous discovery of a previously unknown
quasar at $10.5''$ from Q2225-403 ($z=2.410$).
The observation of the broad emission line of \Mgii{}$_{\lambda 2798}$
and of the surrounding \FeII{} multiplets indicates that the
companion quasar is at $z=0.932$. The spectrum of Q2225-403 shows a number
of absorption lines, the most noteworthy is the \Mgii{} line at the same
redshift of the companion, suggesting that we are probing the gas within
the halo $\sim 80$ kpc from the closer quasar. From high resolution NIR
images we were able to resolve the host galaxies of the two quasars. 
Basing on the known surface density of quasars in the 2dF survey we 
estimate that the probability of finding such a close pair is $\lsim 1$ per cent.
\end{abstract}


\section{Introduction}

Quasar pairs 
can be classified in
physical pairs, gravitational lenses and projected associations. Quasars
in physical pairs are gravitationally interacting or belong to the same
structure (e.g., a cluster of galaxies).
They represent a formidable tool to improve
our understanding of the evolution of galaxy and dark matter
clustering with Cosmic Time, since they can be traced up to very
high redshift \citep{komberg96,shen08}. They can also provide
information about the role of galaxy interactions in triggering
nuclear activity \citep[e.g.,][]{kang07,foreman08}.
Gravitationally lensed quasars allow an unparalleled insight of
the lens distribution of matter \citep[e.g.][]{wittman00,chieregato07}.
Projected pairs can be used as probes of the spatial
structure and ionization properties of intervening intergalactic medium
\citep[e.g.][]{jakobsen86,dodorico08,gallerani08}
and, through the transverse proximity effect \citep{schirber04}, of
the megayear variability and duty cycle of quasars.

Up to now, only a dozen of apparent quasar pairs with angular separation
less than 10 arcsec are known. Recent large field surveys, such as
the Sloan Digital Sky Survey \citep{adelman08}, collected spectra of 
$\sim100,000$ quasars, and probed their large scale ($>0.5$ Mpc) clustering.
Nevertheless, the limit due to the finite physical dimension of the
spectroscopic fibers prevented the observation of objects with angular
separations less than $55''$, making this survey unsuitable for
finding quasar pairs.

\begin{figure*}
\begin{center}
\includegraphics[width=0.46\textwidth]{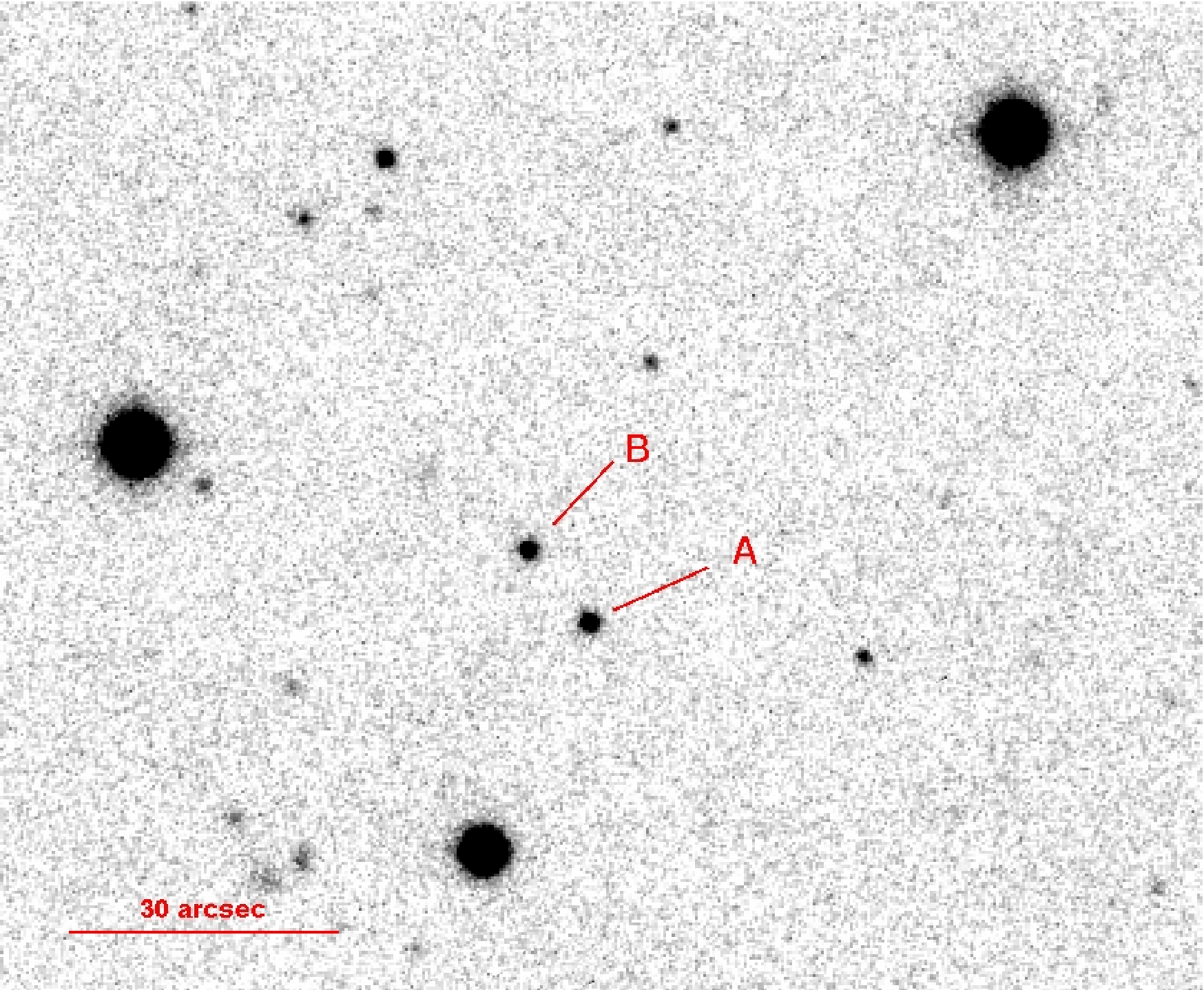}
\includegraphics[width=0.46\textwidth]{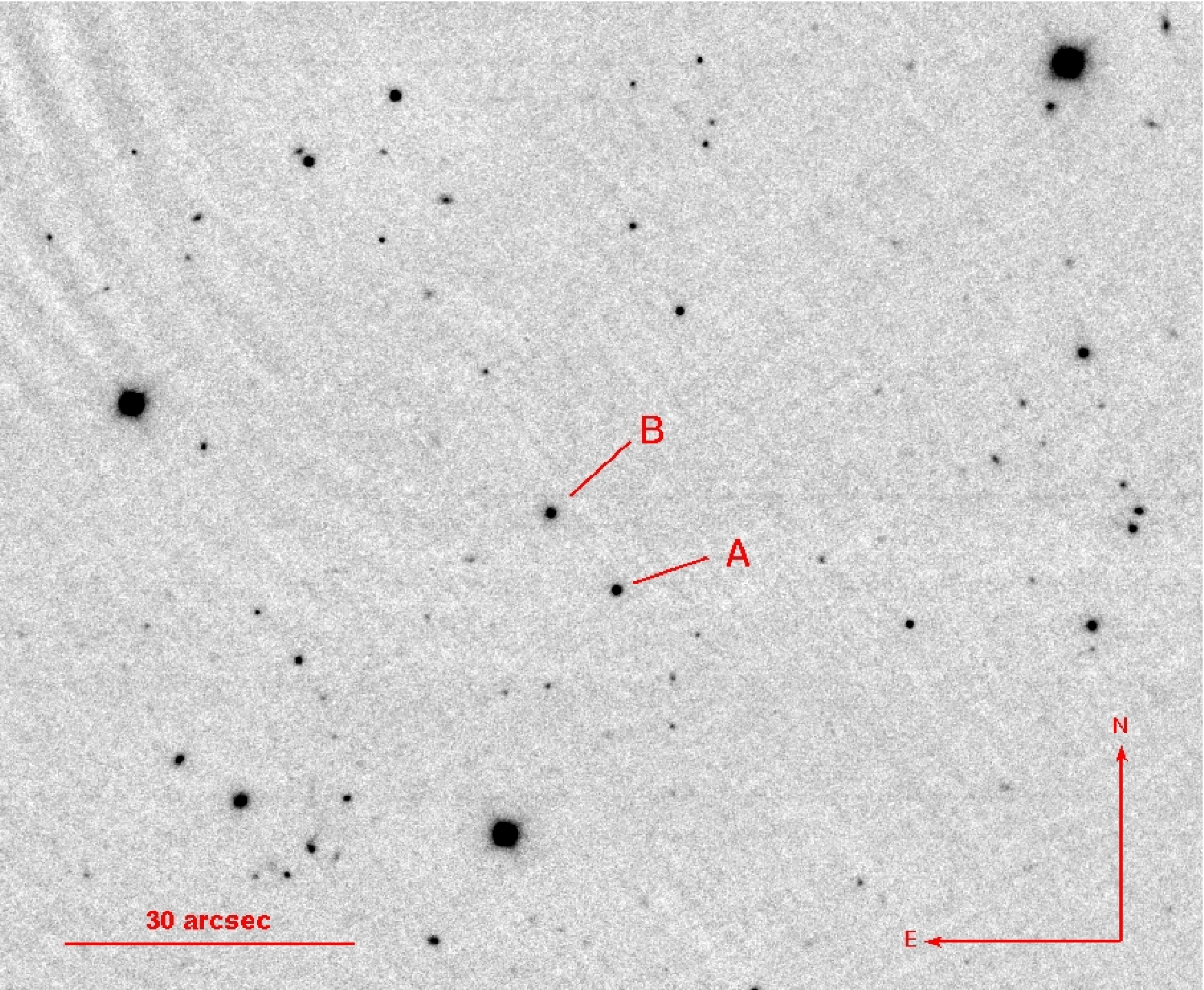}\\
\caption{The field of quasar pair Q2225-403A,B in R (\emph{left}), observed
at the ESO/$3.6$m telescope and in Ks (\emph{right}) taken at the ESO/VLT.
`A' marks the original quasar at $z=2.410$, `B' refers to the new $z=0.932$ quasar.
}\label{fig_q2225_field}
\end{center}
\end{figure*}

In the framework of the study of the \Mbh{}--\Lhost{} relation
throughout Cosmic Time (Decarli et al., 2009b in preparation), we 
collected high-resolution NIR imaging \citep[][K09]{kotilainen09} and 
optical spectroscopy (Decarli et al., 2009a, in preparation) of Q2225-403 
(hereafter, quasar A), a $z=2.410$ quasar first reported by
\citet[][]{hewitt93}.
We set the slit orientation so that we simultaneously
observed both quasar A and the $10.5''$ North-East source with similar
magnitude (see figure \ref{fig_q2225_field}). The spectrum of the 
companion shows it is a quasar at $z=0.932$ (quasar B).
In this paper we discuss the properties of this system together with
a statistical analysis of apparent quasar pairs.

Throughout the paper, we adopt a concordance cosmology
with $H_0=70$ km/s/Mpc, $\Omega_m=0.3$, $\Omega_\Lambda=0.7$.

\section[]{Observations and data reduction}\label{sec_observations}

\subsection[]{Spectroscopic data}
Spectroscopy was collected at the European Southern
Observatory (ESO) $3.6$m telescope in La Silla (Chile; program ID:
079.B-0304(A)) on September, $9^{\rm th}$, 2007: see Decarli et al.
2009a for details. Spectra were obtained through a $1.2''$ slit
in the wavelength $4100$--$7500$ \AA{} with a spectral resolution $R\sim 400$.
Standard recipes for data reduction where adopted. Three
individual exposures, for a total of 71 min integration time, were acquired.
We set the Position Angle to $131.6^\circ$, in order to observe
simultaneously both Q2225-403 A and B. Two-dimensional spectra were bias
subtracted, flat fielded, re-aligned and combined scaling according
to signal to noise. One-dimensional spectra were extracted and
wavelength- and flux-calibrated. The same calibration
procedures were adopted for the two spectra. Absolute flux
calibration was performed using corollary R-band photometry (see
figure \ref{fig_q2225_field}, \emph{left}). Final spectra were then
de-reddened according to the E(B-V) maps from \citet{schlegel98}.

\begin{figure*}
\begin{center}
\includegraphics[width=0.9\textwidth]{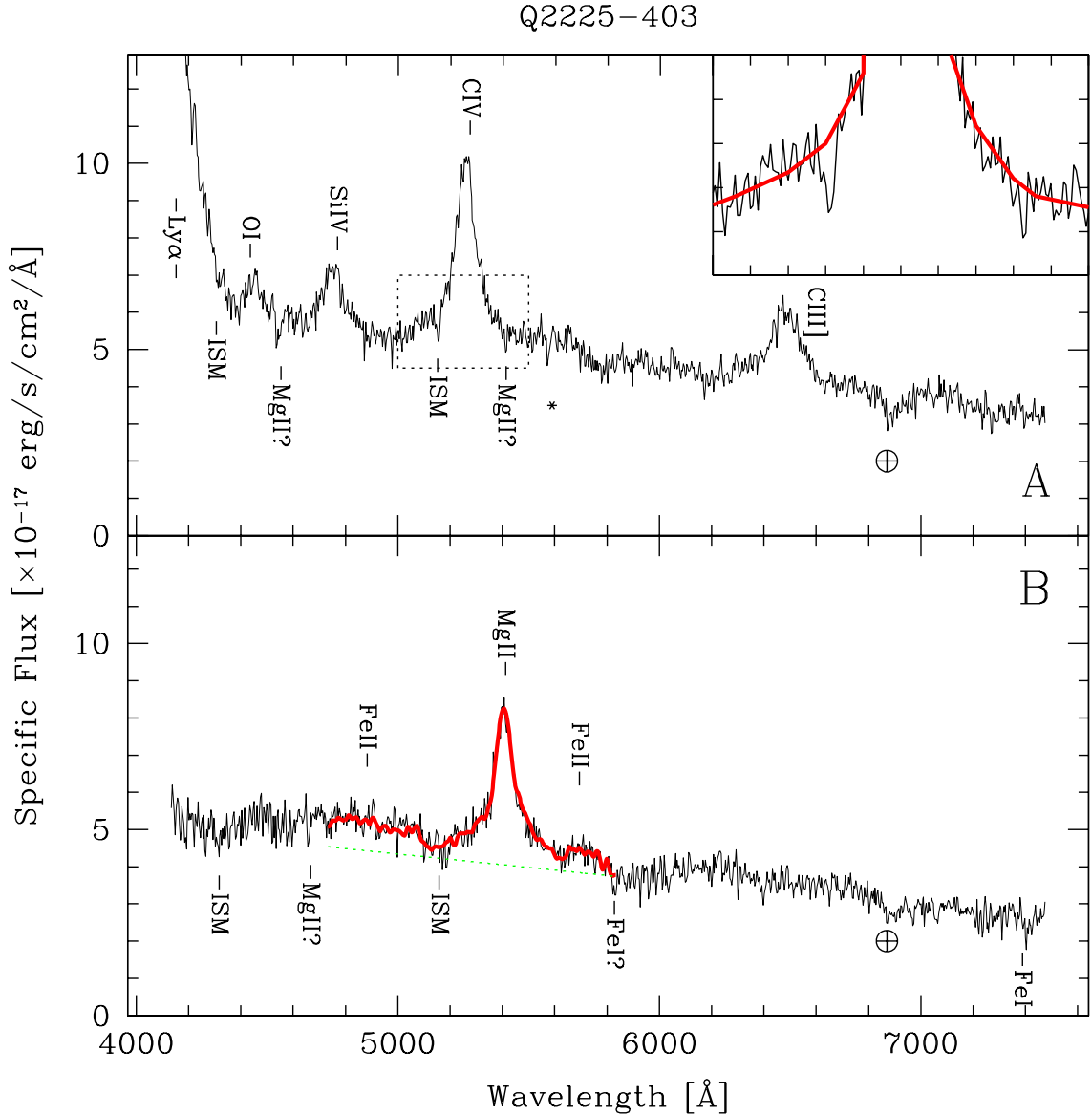}\\
\caption{Spectra of quasars Q2225-403 A and B. Main emission
and absorption lines are labelled (see also table \ref{tab_ewabs}).
The $\oplus$ symbols mark main atmospheric absorptions. The asterisk
refers to the residual of the strong night sky line at 5577 \AA. ISM labels 
the features due to the Galactic interstellar medium. \emph{Top panel -- } 
The inset in the upper
panel highlights the 5000--5500 \AA{} region, with the absorption feature
of \Mgii{} at the bottom of the broad \Civ{} emission 
line. The bold line represent the fit of the \Civ{} emission profile. 
The inset is also marked in the spectrum of quasar A
with a dotted box. \emph{Bottom panel -- } The thick (red) line is the 
model of the MgII+FeII emission \citep{vestergaard01} in the wavelength 
range 4700--5800 \AA{}, supporting the line identification and, hence, 
our estimate of quasar B redshift. The dotted (green) line shows the 
underlying continuum. }\label{fig_spectra}
\end{center}
\end{figure*}

\subsection{Imaging data}

A deep Ks-band image of Q2225-403 field was obtained using ISAAC
\citep{moorwood98}, mounted on UT1 (Antu) of ESO/VLT (see figure
\ref{fig_q2225_field}, \emph{right}). 
Q2225-403 was part of a sample of 16 objects, selected from the
VCV06 catalogue in order to have relatively faint nuclear absolute
magnitudes ($-26>M_{\rm V}>-27$), $2<z<3$ and 2--3 bright stars in
the close field in order to accurately characterize the PSF (this is
mandatory for the study of the host galaxies of bright quasars).
We refer to K09 for details on the observations and data
reduction. The average seeing was $0.46 \pm 0.05$ arcsec and the sky
brightness was $13.34$ mag/arcsec$^2$. Photometric calibration was
performed through the comparison with 2MASS magnitudes of bright
stars available in the field. The estimated photometric
accuracy is $0.05$ mag.

\begin{figure*}
\begin{center}
\includegraphics[width=\textwidth]{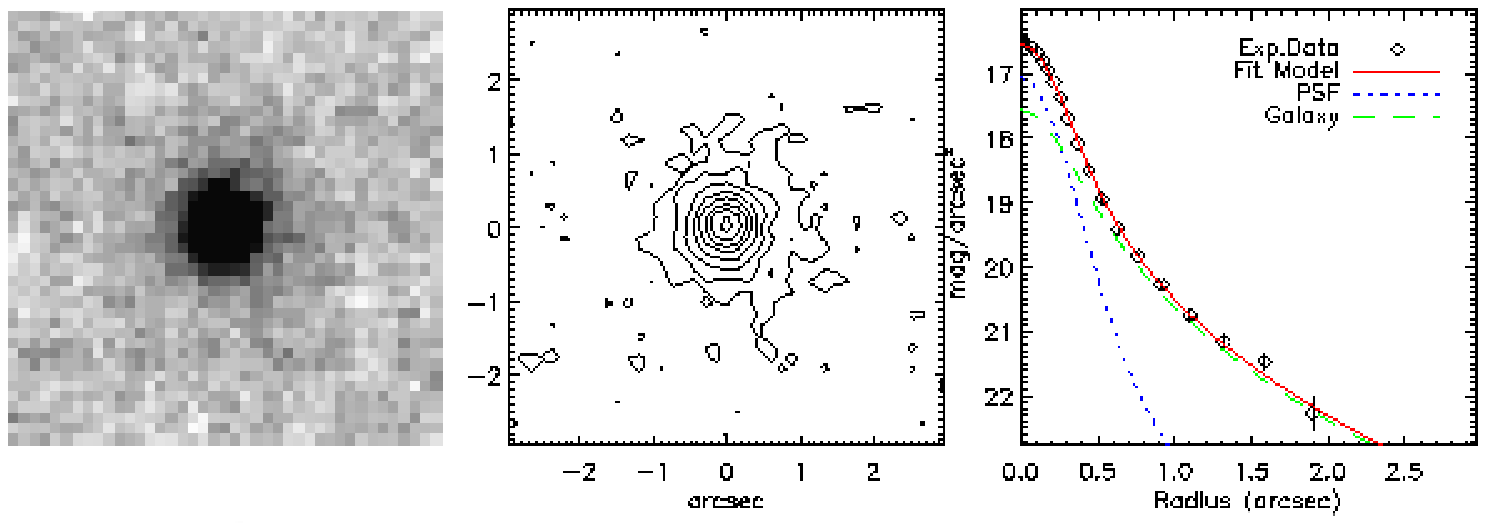}\\
\caption{Analysis of the image of Q2225-403B. \emph{Left} -- the
observed Ks-band image of the quasar; \emph{Centre} -- same image in
isophotal contours; \emph{Right} -- the best-fit light profile model.
See also \texttt{http://www.dfm.uninsubria.it/astro/qso\_host/}.
}\label{fig_aida}
\end{center}
\end{figure*}

\section[]{Results}\label{sec_companion}

\begin{table}
\begin{center}
\caption{List of the main observed absorption features.
Atmospheric and Galactic lines are dropped.
(1) Peak wavelength. (2) Measured Equivalent Width.
(3) Identification. Uncertain classifications are marked with
`?'. (4) Rest frame Equivalent Width, assuming the redshift
in column (3). (5) Spectrum where the lines are detected.} \label{tab_ewabs}
\begin{tabular}{ccccc}
   \hline
$\lambda$&  EW          & Identification         & EW$_r$ &Object\\ 
 \AA     & \AA          &                        & \AA    &     \\ 
 (1)     &   (2)        &  (3)                   & (4)    & (5) \\ 
 \hline
 4547    & $3.8\pm1.4$ & MgII $z=0.625$?	 & $2.3\pm0.9$ & A	\\ 
 5408    & $2.3\pm0.7$ & MgII $z=0.932$?	 & $1.2\pm0.4$ & A	\\ 
\hline
 4661    & $2.3\pm0.6$ & MgII $z=0.665$?         & $1.4\pm0.4$ & B	\\ 
 5833    & $4.4\pm1.0$ & FeI$_{3020}$ $z=0.932$? & $2.3\pm0.5$ & B	\\ 
 7393    & $7.3\pm3.3$ & FeI$_{3830}$ $z=0.932$  & $3.8\pm1.7$ & B	\\ 
   \hline
   \end{tabular}
   \end{center}
\end{table}
Quasar B (RA$_{\rm J2000}$=22:28:50.4; Dec$_{\rm J2000}$=-40:08:27)
is not present in the catalogue by \citet[][hereafter VCV06]{vcv06},
nor in the NED or Simbad archives. It appears in the USNO-B1.0 catalogue
(source ID: 0498-0814048, $m_{\rm R}=20.0$, $m_{\rm B}=20.4$, grossly
consistent with the photometry of our R-band image, $m_{\rm}=19.6$).
The optical spectrum clearly shows the broad \Mgii{} line surrounded by 
\FeII{} multiplets (see Figure \ref{fig_spectra}), leaving no doubt on 
the source nature. The peak of the \Mgii{} line is observed at 5406 \AA{},
yielding $z=0.932$.

A number of absorption systems are apparent in the spectra of
Q2225-403A,B (see Table \ref{tab_ewabs}). In quasar B, an absorption 
at 7393 \AA{} is detected, that is consistent with the
FeI$_{3830}$, typical of early-type galaxies, at $z=0.932$.
Other absorption lines are detected in the spectrum of
quasar A at 4547 \AA{} and B at 4661 \AA{}. Assuming
that they are also produced by \Mgii{}$_{\lambda 2800}$, they
correspond to $z=0.625$ and $z=0.665$ respectively.
Since each feature is observed only in one spectrum, we argue
that the amount of intervening gas of the two clouds drop in a
relatively small spatial scale ($\lsim75$ kpc). Furthermore,
in the spectrum of quasar A a faint feature is present at 5408 \AA{}.
(see Figure \ref{fig_spectra}) This absorption is clearly present 
in each of the 3 individual exposures of the spectrum. Therefore we are 
confident that it is a real feature. The most likely identification of 
this line is with the \Mgii{} doublet ($\lambda\lambda = 2796, 2804$ \AA{}), 
at the same redshift of quasar B, $z=0.932$. At the resolution of our observations,
the two components of the doublet are blended. The presence of this absorption
reveals an extended halo around the host galaxy of quasar B at the projected 
distance of 83 kpc.

From the analysis of NIR imaging data, we are able to detect
the host galaxy of both quasars. For quasar B, the Ks-band
roughly corresponds to the rest-frame J. We convert the observed
magnitude into R-band by assuming the elliptical galaxy template by
\citet{mannucci01}, yielding $M_{\rm R}(\rm host)=-23.2$. The galaxy 
is well resolved and modelled with a de
Vaucouleurs profile with scale radius $R_{\rm eff}=5.2$ kpc (see
Figure \ref{fig_aida}). The nuclear component has $M_{\rm R} (\rm
nuc) = -22.5$. Assuming the quasar template in \citet{francis91}, we
infer R-$i$=$-0.04$. The derived $i$-band nuclear magnitude places
source B at the faint end of quasar luminosity function,
but matching the usual $M_i<-22$ criterion for quasar
classification \citep[e.g.][]{richards06}.

In a study of quasar--galaxy projected associations, \citet{kacprzak07}
found a correlation between the EW of \Mgii{}$_{\lambda2796}$ and the asymmetry of
the galaxy, suggesting a connection between the intervening metal absorption 
systems and the properties of the galaxy environment. According to this
relationship, the stronger are the absorptions, the more disturbed
is the morphology of the galaxy. In our case, we are unable to resolve
the \Mgii$_{\lambda2796}$ line. If we assume a ratio between the 
two components of the \Mgii{} doublet of $1.7$, we infer \Mgii$_{\lambda2796}$=0.8. 
We note that a weak indication of asymmetry is apparent in the faintest
surface brightness level of the host galaxy of quasar B, that is in qualitative
agreement with the trend suggested by \citet{kacprzak07}.

\begin{table*}
\begin{center}
\caption{Quasar apparent pairs with $\theta<10''$ from the VCV06
catalogue, in a comparison with Q2225-403A,B.
(1--4) $z$, RA, Dec and $m_{\rm V}$ of the quasar with the
lower $z$. (5--8) the same for the higher $z$ quasar.
(9) angular separation. (10) projected separation assuming
the redshift of the lower-$z$ quasar. (11) redshift difference.
 (12) references (only those references concerning quasar pairs 
 are included):{\it a} - \citet[][]{burbidge97}; {\it b} - 
 \citet[][]{sluse03}; {\it c} - \citet[][]{hennawi06}; {\it d} - 
 \citet[][]{myers07}.} \label{tab_vcvpairs}
\begin{tabular}{cccc|cccc|cccc}
   \hline
      \multicolumn{4}{c}{Nearer quasar}       & \multicolumn{4}{c}{Farther quasar}  &$\theta$&Proj.sep.&$\Delta z$&Ref\\
 $z$  & RA(J2000) & Dec(J2000) &$m_{\rm V}$&$z$& RA(J2000)  & Dec(J2000) &$m_{\rm V}$ & [''] & [kpc] &  &      \\
 (1)  &   (2)     &  (3)       &  (4)  &  (5)  &   (6)      &  (7)      & (8)   &  (9) &  (10) &(11)    & (12)  \\
 \hline
\multicolumn{11}{l}{\qquad{}\emph{Known quasar pairs}} \\
1.545 & 00:02:12.6 & -00:53:11 & 20.46 & 2.206 & 00:02:12.1 & -00:53:09 & 19.58 &  6.3 & 53 & 0.661 & {\it c}   \\
2.030 & 00:40:18.2 & +00:55:31 & 18.72 & 2.086 & 00:40:18.7 & +00:55:26 & 19.67 &  7.7 & 64 & 0.056 & {\it c}   \\
1.296 & 01:22:12.7 & +14:10:54 & 19.91 & 1.579 & 01:22:13.1 & +14:10:52 & 20.03 &  8.1 & 67 & 0.283 & {\it c}   \\
1.442 & 02:41:06.9 & +00:10:27 & 19.89 & 1.673 & 02:41:07.4 & +00:10:28 & 20.87 &  9.8 & 82 & 0.231 & {\it c}   \\
2.180 & 08:14:20.4 & +32:50:16 & 19.86 & 2.210 & 08:14:19.6 & +32:50:19 & 20.33 &  9.5 & 78 & 0.030 & {\it c}   \\
1.340 & 09:02:35.7 & +56:37:56 & 20.30 & 1.390 & 09:02:35.4 & +56:37:51 & 20.63 &  6.6 & 55 & 0.050 & {\it c,d} \\
1.627 & 10:12:15.8 & -03:07:08 & 18.90 & 2.746 & 10:12:15.8 & -03:07:03 & 17.60 &  4.5 & 38 & 1.119 & {\it a,b} \\  
1.142 & 12:04:50.5 & +44:28:35 & 19.20 & 1.814 & 12:04:50.7 & +44:28:33 & 19.42 &  3.8 & 31 & 0.672 & {\it c,d} \\
2.379 & 12:25:45.7 & +56:44:40 & 19.36 & 2.390 & 12:25:45.2 & +56:44:45 & 20.51 &  7.4 & 60 & 0.011 & {\it c}   \\
2.001 & 12:49:48.1 & +06:07:08 & 20.42 & 2.376 & 12:49:48.2 & +06:07:13 & 20.37 &  3.9 & 32 & 0.375 & {\it c,d} \\
0.436 & 15:50:43.7 & +11:20:47 & 17.23 & 1.901 & 15:50:44.0 & +11:20:47 & 18.78 &  4.4 & 24 & 1.465 & {\it a,b} \\
\hline
\multicolumn{11}{l}{\qquad{}\emph{New quasar pairs}} \\
1.310 & 00:39:54.3 & -27:25:23 & 20.61 & 2.100 & 00:39:54.1 & -27:25:14 & 20.64 &  9.9 & 83 & 0.790 &       \\
1.264 & 00:39:54.8 & -27:25:20 & 20.26 & 1.310 & 00:39:54.3 & -27:25:23 & 20.61 &  4.3 & 35 & 0.046 &        \\ 
1.333 & 01:10:50.8 & -27:19:51 & 20.10 & 2.261 & 01:10:51.4 & -27:19:57 & 20.84 &  9.3 & 78 & 0.928 &         \\
1.586 & 03:42:12.4 & -44:16:41 & 19.20 & 2.077 & 03:42:12.6 & -44:16:36 & 19.70 &  5.7 & 48 & 0.491 &        \\
1.519 & 10:16:36.3 & -02:34:22 & 19.55 & 2.617 & 10:16:36.4 & -02:34:12 & 20.31 &  8.9 & 75 & 1.098 &     \\
0.649 & 10:51:26.8 & -02:27:20 & 19.43 & 1.160 & 10:51:26.3 & -02:27:17 & 19.53 &  8.0 & 55 & 0.511 &      \\
0.888 & 11:18:47.9 & +40:26:43 & 20.70 & 1.129 & 11:18:48.6 & +40:26:47 & 19.05 &  7.9 & 61 & 0.241 &      \\  
1.863 & 23:33:05.3 & -28:00:54 & 20.33 & 1.970 & 23:33:04.7 & -28:00:55 & 19.72 &  7.5 & 63 & 0.107 &      \\
   \hline
\multicolumn{11}{l}{\qquad{}\emph{This paper}} \\
0.932 & 22:28:50.4 & -40:08:27 & 20.20 & 2.410 & 22:28:49.9 & -40:08:34 & 20.10 & 10.5 & 83 & 1.478 &      \\
   \hline
   \end{tabular}
   \end{center}
\end{table*}

\section{An inventory of apparent quasar pairs}\label{sec_probability}

Due to the importance of quasar pairs of the type discussed here, 
we made an inventory of similar systems starting from the VCV06 catalogue.
Out of $\sim85000$ quasars, we found 19
pairs with angular separation $\theta<10''$ and line-of-sight velocity
differences exceeding 3000 km/s (hence excluding physical pairs). They are
listed in Table \ref{tab_vcvpairs}. We note that only 11 systems out of 19 
have been already considered in the framework of quasar pairs. 
On average, apparent pairs reported in Table \ref{tab_vcvpairs} have
$<$$z(\rm near)$$>\approx1.4$ and $<$$\Delta z$$>\approx0.5$.
Our case represents a record in terms of redshift difference.
Other three systems are reported with the nearer quasar at $z<1$, where
a detailed study of the host galaxy luminosity and morphology is feasible.
The typical projected distances at $z(\rm near)$ are $\sim 60$ kpc.

A number of apparent quasar pairs have been proposed as anomalous 
associations \citep[e.g.][]{burbidge97,galianni05} with respect to chance 
alignments. We estimate that the number of systems reported in Table 
\ref{tab_vcvpairs} is consistent 
with the assumption of chance superposition. In fact, the probability that, 
given a quasar, a projected 
companion can be found within a given angular separation $\theta$
follows the Poisson statistics: $P(<\theta)\approx \lambda$.
Here $\lambda$ is the expected number of quasars 
in the solid angle defined by $\theta$, $\lambda=\rho(<$$m)\, \pi \theta^2$, 
and $\rho\,(<$$m)$ is the number density of quasars brighter than a given 
magnitude $m$. We refer to the 2dF survey \citep{croom04}: $\rho\,(m_b$$<$$20)=13.8$ 
quasars per square degree, in good agreement with the values from the SDSS
\citep[see][]{yanny00}.
%
%
Hence the probability of finding a quasar with $m_b\lsim20$
within a $10.5''$ circle is $\sim4\times10^{-4}$.

\section{Conclusions}\label{sec_conclusions}

We report the discovery of an apparent quasar pair with angular
separation of $10.5''$.  Q2225-403A,B is the only apparent pair of 
quasars for which both the 
host galaxies have been resolved. Their Ks-band apparent 
magnitudes ($m_{\rm A}(\rm host) = 18.51$; $m_{\rm B}(\rm host) = 17.44$)
are consistent with those expected for typical quasar host galaxies 
at the distance indicated by their redshift. The discovery of 
an intervening absorption system in quasar A at the same redshift of 
B reveals an extended halo around the nearest object. 

Based on the known surface density distribution
of quasars we find that the \emph{a priori} probability of finding
such a pair in our survey is of the order of 0.6 per cent and
the discovery of the absorption system on the spectrum of quasar A
at the same $z$ of quasar B is a clear evidence that these two objects
are an apparent pair.

We propose a list of apparent quasar pairs which deserve a specific study
to investigate the properties of the extended halo around quasar host galaxies.

\section*{Acknowledgments}

We thank Marzia Labita for useful discussions. This work was based on
observations made with the ESO/3.6m telescope
in La Silla and with the ESO/VLT in Paranal. This research has made
use of the \emph{VizieR Service}, available
at \texttt{http://vizier.u-strasbg.fr/viz-bin/VizieR} and of the
NASA/IPAC Extragalactic Database (NED) which is operated by the Jet
Propulsion Laboratory, California Institute of Technology, under
contract with the National Aeronautics and Space Administration.

\label{lastpage}

\bsp
\end{document}